%% file: main.tex
\pdfoutput=1
\documentclass[aps,prd,amsmath,floats,floatfix, twocolumn,
superscriptaddress,nofootinbib,showpacs,longbibliography]{revtex4-1}
\input{preamble}
\usepackage{orcidlink}
\usepackage{enumitem}
\DeclareMathAlphabet{\mathpzc}{OT1}{pzc}{m}{it}

\newcommand{\roughly}{\mathchar"5218\relax\,} 

\newcommand{\h}{\mathpzc{h}}

\newcommand{\dd}{\text{d}}

\newcommand{\sFt}{\widetilde{F}^{\text{S}}(t)\xspace}
\newcommand{\sFtt}{\widetilde{\mathbb{F}}^{\text{S}}(t)\xspace}
\newcommand{\Ft}{\widetilde{F}(t)\xspace}
\newcommand{\Ftt}{\widetilde{\mathbb{F}}(t)\xspace}
\newcommand{\Ff}{F(f)\xspace}
\newcommand{\sFf}{F^{\text{S}}(f)\xspace}
\newcommand{\hf}{\mathpzc{h^{\text{UL}}(f)}}
\newcommand{\sLhf}{\mathpzc{h^{\text{SL}}(f)}}

\begin{document}
\title{Surrogate modeling of gravitational waves microlensed by spherically symmetric potentials}

\newcommand{\ICTS}{\affiliation{International Centre for Theoretical Sciences, Tata Institute of Fundamental Research, Bangalore 560089, India} }
\newcommand{\IUCAA}{\affiliation{Inter-University Centre for Astronomy and Astrophysics, Post Bag 4, Ganeshkhind, Pune 411007, India}}
\newcommand{\SNU}{\affiliation{Department of Physics and Astronomy, Seoul National University, Seoul 08826, Korea} }
\newcommand{\VSM}{\affiliation{Department of Physics, Vivekananda Satavarshiki Mahavidyalaya (affiliated to Vidyasagar University), Manikpara 721513, West Bengal, India} }
\newcommand{\UMD}{\affiliation{Department of Mathematics, Center for Scientific Computing and Visualization Research, University of Massachusetts Dartmouth, Massachusetts 02747, USA} }

\author{Uddeepta Deka\,\orcidlink{0000-0002-5942-4487}}
\email{uddeepta.deka@icts.res.in}
\ICTS
\author{Gopalkrishna Prabhu\,\orcidlink{0009-0001-2695-3622}}
\email{gopal.prabhu@iucaa.in}
\IUCAA
\author{Md Arif Shaikh\,\orcidlink{0000-0003-0826-6164}}
\email{arifshaikh.astro@gmail.com}
\VSM
\SNU
\author{Shasvath J. Kapadia\,\orcidlink{0000-0001-5318-1253}}
\email{shasvath.kapadia@iucaa.in}
\IUCAA
\author{Vijay Varma\,\orcidlink{0000-0002-9994-1761}}
\email{vijay.varma392@gmail.com}
\UMD
\author{Scott E. Field\,\orcidlink{0000-0002-6037-3277}}
\email{sfield@umassd.edu}
\UMD

\hypersetup{pdfauthor={Deka et al.}}


\date{\today} 

\begin{abstract}
The anticipated observation of the gravitational microlensing of gravitational
waves (GWs) promises to shed light on a host of astrophysical and cosmological
questions. However, extracting the parameters of the lens from the modulated
GWs requires accurate modeling of the lensing amplification factor, accounting
for wave-optics effects. Analytic solutions to the lens equation have not been
found to date, except for a handful of simplistic lens models. While numerical
solutions to this equation have been developed, the time and computational
resources required to evaluate the amplification factor numerically make large-scale parameter estimation of the lens (and source) parameters prohibitive. On
the other hand, surrogate modeling of GWs has proven to be a powerful tool to
accurately, and rapidly, produce GW templates at arbitrary points in parameter
space, interpolating from a finite set of available waveforms at discrete
parameter values. In this work, we demonstrate that surrogate
modeling can also effectively be applied to the evaluation of the time-domain
microlensing amplification factor $\widetilde{F}(t)$. We show this by constructing $\widetilde{F}(t)$ for
two lens models, viz. point-mass lens, and singular isothermal sphere, which notably includes logarithmic divergence behaviour. We find both surrogates reproduce the original lens models accurately, with mismatches $\lesssim 5 \times 10^{-4}$ across a range of plausible microlensed binary black hole sources observed by the Einstein Telescope. This surrogate is between 5 and $10^3$ times faster than the underlying lensing models, and can be evaluated in about 100 ms. The accuracy and efficiency attained by our surrogate models will enable practical parameter estimation analyses
of microlensed GWs.

\end{abstract}


\maketitle

\section{Introduction}
\label{sec:introduction}

The LIGO-Virgo-KAGRA (LVK) gravitational-wave (GW) detector network
~\cite{TheLIGOScientific:2014jea, TheVirgo:2014hva, KAGRA:2020tym} has detected $\sim 90$
compact binary coalescence (CBC) events across three observing runs (O1, O2,
O3)~\cite{LIGOScientific:2018mvr, LIGOScientific:2020ibl, LIGOScientific:2021usb, KAGRA:2021vkt}. The overwhelming majority of these are stellar-mass
binary black holes (BBHs), although binary neutron star (BNS)
~\cite{TheLIGOScientific:2017qsa, Abbott:2020uma} and neutron star black hole
(NSBH) mergers have also been observed ~\cite{LIGOScientific:2021qlt}.

These observations have enabled tests of general relativity (GR) in the
strong-field regime \cite{LIGOScientific:2021sio}, provided distance-ladder
independent measurements of the Hubble constant \cite{LIGOScientific:2021aug},
given hints of the population properties of BBHs \cite{KAGRA:2021duu}, and
allowed probes of matter at densities not accessible in current human-made
laboratories via constraints on the NS equation of state
\cite{LIGOScientific:2018cki}. The ongoing 
observing run (O4) will likely triple the number of observed events, thus further strengthening the
astrophysical and cosmological inferences made from O1-O3. Moreover, the
increased detections will also improve our chances of seeing a gravitationally
lensed GW event, which, to date, has not been detected
\cite{LIGOScientific:2023bwz}.

Gravitational lensing of GWs, akin to that of light, occurs when they encounter
agglomerations of matter in their paths. However, unlike the lensing of
electromagnetic (EM) waves, that of GWs can be categorized into two regimes. If
the wavelength of the GWs ($\lambda_{\mathrm{GW}}$) is significantly smaller
than the gravitational radius of the lens ($\lambda_{\mathrm{GW}} \ll
GM_{\mathrm{L}}/c^2$)\footnote{$M_{\rm L}$ is the mass of the lens, $G$ is the Newton's gravitational constant and $c$ is the speed of light in vacuum.}, then two or more temporally resolved GW images will
be formed, with identical frequency evolution as the source, but differing
amplitudes \cite{wang1996, Dai:2017huk, ezquiaga2021}. The time delay of the
images could range from minutes to months for galaxy-scale lenses \cite{Ng2018,
li2018gravitational, oguri2018effect}, and weeks to years for cluster-scale
lenses \cite{smith2018if, smith2020massively,Smith:2019qsv, robertson2020does,
ryczanowski2020building}. Geometric optics is a good approximation in such cases. Conversely, if $\lambda_{\mathrm{GW}} \sim
GM_{\mathrm{L}}/c^2$, a single modulated image with wave optics effects
(such as beating and diffraction patterns) will be produced
\cite{deguchi1986,nakamura1998, takahashi2003wave, cao2014, lai2018,
christian2018, dai2018, diego2020, Yeung:2021chy, Seo:2021psp, Wright:2021cbn}\footnote{Note that there could also be wave-optics effects due to interference in certain lens configurations even when $\lambda_{\rm GW} \ll GM_{\rm L}/c^2$ (see, e.g., ~\cite{Liu:2023ikc}).}. Lensing promises to significantly enhance
the science that can be done with GWs, such as enhanced GW early-warning of
electromagnetically bright CBCs \cite{magare2023gear}, additional
unique tests of GR \cite{baker2017, collett2017, fan2017, goyal2021,
ezquiaga2020}, precisely measuring the expansion rate of the universe
\cite{sereno2011cosmography,liao2017precision, cao2019direct,
li2019constraining, hannuksela2020localizing, jana2023}, enable constraints on
the structure of the lens (including probing properties such as electric
charge) \cite{Deka:2024ecp}, as well as on the fraction of dark matter in the form
of massive compact objects \cite{Basak:2021ten, Barsode:2024wda}.

Gravitational microlensing of the LVK's GWs falls in the wave-optics
regime. Examples of corresponding microlenses include (but are not limited to) isolated intermediate-mass black holes (IMBHs), whose masses lie in the range $M \in \sim [10^2,
10^4] M_{\odot}$. 
Currently, searches for microlensing/wave-optics patterns
(see, e.g., \cite{LIGOScientific:2023bwz}) in detected CBC events assume a
point-mass lens. This assumption is made for two reasons. The first is that an
exact, analytical form of the frequency domain microlensing amplification
factor, $F(f)$, is known (see, e.g, \cite{takahashi2003wave}), which facilitates the creation of an interpolation table for $F(f)$. The evaluation
of this 
interpolant, for arbitrary GW frequencies and lens parameters
(viz., lens mass and impact parameter), is sufficiently rapid to attempt large-scale parameter estimation runs without prohibitively taxing computational
resources \cite{hannuksela2019search}. The second is that IMBHs as microlenses
have been argued to be well-modelled by the point-mass lens \cite{lai2018}.

However, other models of microlenses exist, many of which do not currently have
any known analytical form for $F(f)$. An example of particular astrophysical
relevance is microlens(es) embedded in a macro potential. It has been argued
that a microlens, such as a massive compact object, lying in the halo of an
intervening galaxy that provides the macro-potential, could result in the
production of resolvable images with wave-optics effects imprinted on each of
them \cite{Meena:2019ate}. Moreover, multiple microlenses in the macropotential
would additionally produce interference patterns between the images
\cite{diego2019observational, pagano2020lensinggw, cheung2021stellar,
Mishra:2021xzz}. The resulting amplification factor $F(f)$, containing beating,
diffraction and interference patterns, has no known analytical form to date.

Constructing such realistic $F(f)$'s requires solving the lens equation, with
the appropriate superposition of lensing potentials, numerically. This has been
achieved (see, e.g., \cite{Mishra:2021xzz}) using the method described in
\cite{ulmer1995}. However, producing these $F(f)$'s, for a single set of lens
parameters, typically takes several seconds or longer per waveform, making large-scale GW parameter estimation (PE) of such lens configurations unfeasible. 

In this work, we adopt a widely used interpolation technique, called ``surrogate modeling''. This technique has found its application in the field of GWs for modeling waveforms in the
context of CBCs
~\cite{Field:2013cfa,Blackman_2017,Blackman:2017pcm,Varma:2019csw,Islam:2021mha,Yoo:2022erv,Yoo:2023spi,Gerosa:2018qay,Varma:2018aht,Islam:2023mob,rink2024gravitationalwavesurrogatemodel, Islam_2022}; see Sec.~5 of Ref.~\cite{LISAConsortiumWaveformWorkingGroup:2023arg} for a summary of recent CBC applications. In addition, a recent study has demonstrated the application of surrogate modeling to describe waveforms for hyperbolic encounters between black holes ~\cite{fontbuté2024numericalrelativitysurrogatemodelhyperbolic}. The
main appeal of surrogate modeling lies in the fact that one can obtain a fast
and accurate prediction of a high-dimensional, complicated function at
arbitrary points in parameter space, interpolating from a finite (and usually
small) set of points in that space where the function is known. To achieve
this, surrogate models take advantage of the underlying similarity among the
numerical solutions across the parameter space. Surrogate models thus provide a
drastic increase in the evaluation speed of the function, compared to slow
numerical solutions, while only negligibly deviating from them. We apply surrogate modelling, {\it for the first time}, to the rapid and accurate construction of
time domain lensing amplification factors $\widetilde{F}(t)$. We benchmark the accuracy and production speeds of the surrogate microlensed
GWs in the frequency domain. 

As a proof of principle, we consider the following lens models: point-mass lens and singular isothermal sphere (SIS). We numerically evaluate the time domain
amplification factor $\widetilde{F}(t)$ at a few discrete points in the lensing
parameter space. From these, we construct surrogate amplification factors
$\widetilde{F}^{S}(t)$ for each of the lensing configurations. After Fourier
transforming and producing microlensed GW waveforms in the frequency domain, we
evaluate the production-time and accuracy of the surrogate microlensed
waveforms. We find that we're able to achieve mismatches of the order of
$\mathcal{O}(10^{-7} - 10^{-3})$ with respect to the waveforms evaluated
numerically, with evaluation times of $\mathcal{O}(10^{-1} - 10^{-2})$s. These
benchmarking tests showcase the efficacy of surrogate modeling applied to
microlensed GWs. They also suggest that surrogate modeling can be feasibly used
for GW parameter estimation in the context of GW microlensing.

The rest of the paper is organised as follows. Section~\ref{sec:method} delineates the lensing wave-optics-driven amplification factor for the two lens models considered in this work. It also introduces the basics of surrogate modelling and its application to the construction of time-domain lensing amplification factors. Section~\ref{sec:results} presents the results which demonstrate the accuracy and evaluation speed of the surrogate microlensed waveforms. Section~\ref{sec:conclusions} summarises the results and suggests future work. Details about the lens models used in this work are described in Appendix \ref{appx-a}.

\section{Method}
\label{sec:method}
In this section we outline the key steps in constructing the surrogate microlensed waveforms. We begin by introducing the lensing amplification factor in Section~\ref{sec:lensing_amplification_factor}, which is the quantity capturing the effect of microlensing on unlensed waveforms. This is followed by the description of the peak reconstruction procedure in Section~\ref{sec:peak_reconstruction}, an essential step to accurately model the contribution of the region in the lens plane near saddle images to the time domain amplification factor. We then introduce an amplitude regularization technique in Section~\ref{sec:time_domain_ampl_transf} to assist the surrogate to better interpolate between waveforms. In Section~\ref{sec:surrogate_modeling} we present the details of the surrogate modeling approach.

\subsection{Lensing Amplification Factor}
\label{sec:lensing_amplification_factor}
\begin{figure}[h]
      \centering
      \includegraphics[width=0.48\textwidth]{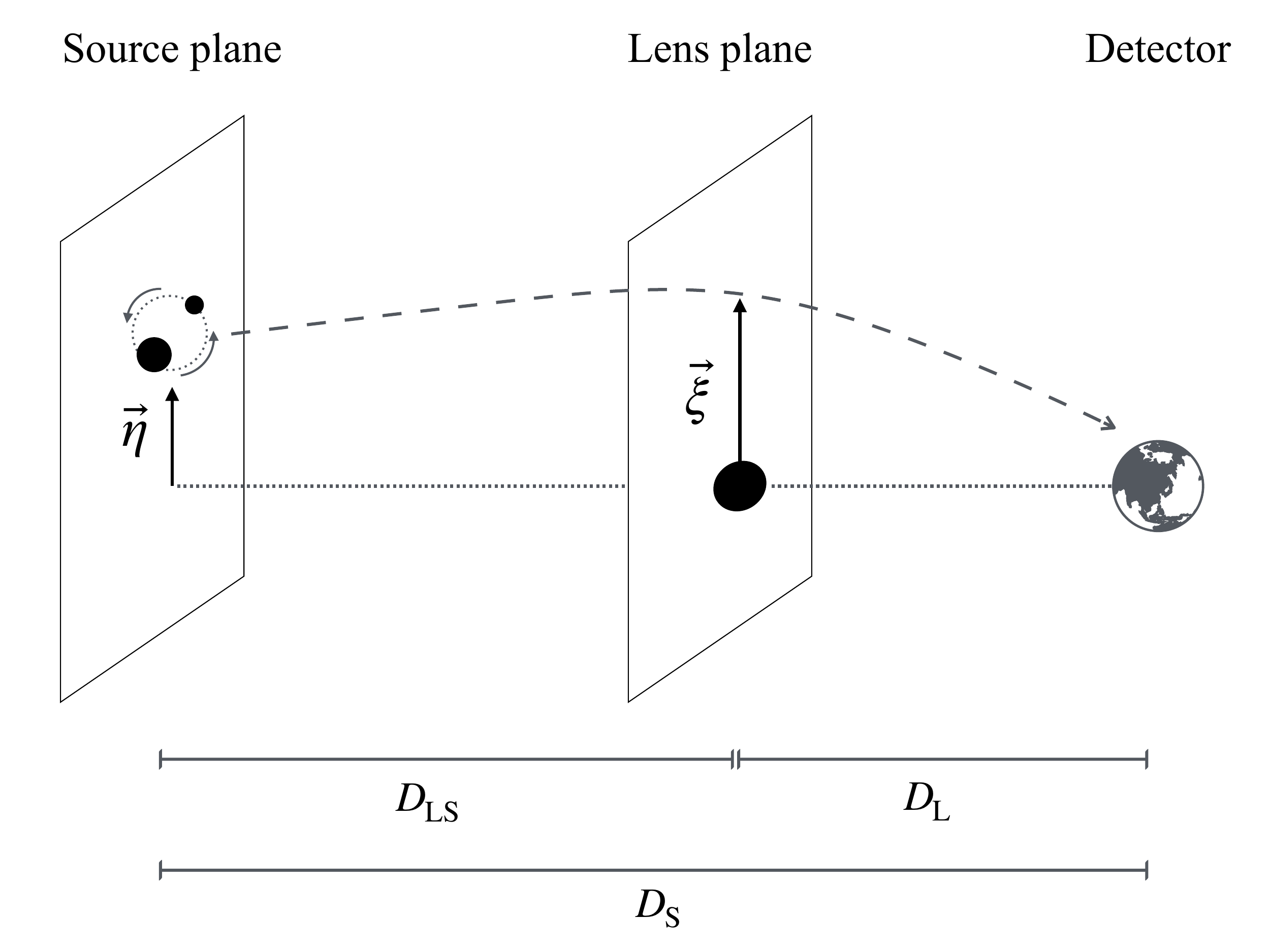}
      \caption{Lensing geometry for the source (a compact binary merger) at redshift $z_{\rm S}$, the lens (compact object within the thin lens approximation) at redshift $z_{\rm L}$, and the ground-based detector. $\vec{\xi}$ denotes the lens plane coordinates. The source is located at $\vec{\eta}$ measured from the optical axis (dotted line). $D_{\rm L}, D_{\rm S} $ and $D_{\rm LS}$ are the angular diameter distances between the detector and the lens, between the detector and the source, and between the lens and the source, respectively. The dashed line denotes the deflected path of the incoming signal.}
      \label{fig:lens_config}
\end{figure}

In this section, we describe the formalism for computing the lensing amplification factor $F(f)$, which is defined as the ratio between the lensed and the unlensed GW waveforms,
\begin{equation}
    F(f) \equiv \mathpzc{h^{\text{L}}(f)}/\hf ~.    
\end{equation}
Under the thin lens approximation, the frequency-domain amplification factor is given by the Kirchhoff diffraction integral \cite{schneider2013gravitational}:
\begin{equation}
    \label{eq:ampfac_fd}
    F(f)=\frac{D_{\rm S}\xi_{0}^{2}(1+z_{\rm L})}{cD_{\rm L}D_{\rm LS}}\frac{f}{i}\int \dd^{2}\vec{x}\exp\left[2\pi if t_{\rm d}(\vec{x},\vec{y})\right]~,
\end{equation}
where $\xi_{0}$ is a typical length scale in the lens plane, $z_{\rm L}$ is the lens redshift, $D_{\rm L}$ is the angular diameter distance to the lens, $D_{\rm S}$ is the angular diameter distance to the source and $D_{\rm LS}$ is the angular diameter distance between the source and the lens (see Fig.~\ref{fig:lens_config} for an illustration). 
The integral is over the dimensionless length vector on the lens plane $\vec{x}\equiv \vec{\xi}/\xi_{0}$. Similarly, $\vec{y}\equiv(D_{\rm L}/\xi_{0}D_{\rm S})\vec{\eta}$ is the dimensionless impact parameter vector on the source plane indicating the position of the source. The time delay function, $t_{\rm d}$, as measured by an observer, is a combination of $t_{\rm geom}$ and $t_{\rm grav}$, where the former is the time delay due to the extra path length of the deflected GW signal relative to an unperturbed null geodesic and the latter is the time delay due to gravitational time dilation:
\begin{align}
  \label{eq:time_delay}
  t_{\rm d} (\vec{x},\vec{y}) &\equiv t_{\rm geom} + t_{\rm grav} =\frac{D_{\rm S}\xi_{0}^{2}\left(1+z_{\rm L}\right)}{cD_{\rm L}D_{\rm LS}}\left[\frac{|\vec{x}-\vec{y}|^{2}}{2}-\psi(\vec{x}) \right]~.
\end{align}
Here, the lensing potential $\psi(\vec{x})$ is the appropriately scaled, projected Newtonian potential on the two-dimensional lens plane that depends on the geometry of the lens. Further, it is convenient to introduce the dimensionless angular frequency $w$ defined as:
\begin{equation}
  \label{eq:dimensionless_frequency}
  w \equiv \frac{D_{\rm S}}{cD_{\rm L}D_{\rm LS}}\xi_0^2(1+z_{\rm L})2\pi f ~.
\end{equation}
The amplification factor in eq.\eqref{eq:ampfac_fd} then becomes:
\begin{equation}
  \label{eq:ampfac_dimensionless_fd}
  F(w) = \frac{w}{2\pi i}\int \dd^{2}\vec{x}\;\exp\left[iw\tau_{\rm d}(\vec{x},\vec{y})\right],
\end{equation}
where $\tau_{\rm d}$ is the dimensionless time delay, given by:
\begin{equation}
  \label{eq:time_delay_dimensionless}
  \tau_{\rm d}(\vec{x},\vec{y}) = \frac{|\vec{x}-\vec{y}|^{2}}{2}-\psi(\vec{x}).
\end{equation}

The integrand in Eq. \eqref{eq:ampfac_dimensionless_fd} is highly oscillatory for large values of $w\tau_{\rm d}$. With the exception of the point-mass lens (PML) model, for which there exists a closed-form analytic $F(w)$, it is computationally expensive to solve the integral using conventional numerical integration techniques for other astrophysically relevant lens models. Moreover, it can be difficult to model oscillatory functions using surrogates. Therefore, we compute the amplification factor in the time domain, similar to the idea proposed in \cite{ulmer1995}. For this purpose we define $\widetilde{F}(t)$ as the inverse Fourier transform of $\left[2\pi i F(w)/w\right]$, which gives:
\begin{equation}
  \widetilde{F}(t) =\int \dd^2\vec{x}\;\delta\left[\tau_{\rm d}(\vec{x}, \vec{y})-t\right] = \frac{\dd S}{\dd t},\label{eq:ampfac_td}
\end{equation}
where we identify that the contribution to $\widetilde{F}(t)$ comes from the area $dS$ on the lens plane, between the curves of constant time delays $t$ and $t+dt$. Therefore, the computation of $\widetilde{F}(t)$ boils down to finding areas between nearby time delay curves, which is done using the method described in \cite{Deka:2024ecp}. 

We can then obtain $F(w)$ by Fourier transforming $\widetilde{F}(t)$ as shown below:
\begin{equation}
  \label{eq:FwFFT}
  F(w) = \frac{w}{2\pi i}\times\rm{FFT}\left[\widetilde{F}(t)\right]~,
\end{equation}
where $\text{FFT}[\cdot]$ refers to the fast Fourier transform routine. Finally, we can go from the dimensionless angular frequency $w$ to the dimension-full frequency $f$ by setting the length scale $\xi_0$ appropriately. Throughout this work, we set $\xi_0$ to be the \textit{Einstein radius} of the lens (see Eq. \eqref{eq:einstein_radius}), simplifying Eq.~\eqref{eq:time_delay} and Eq.~\eqref{eq:dimensionless_frequency} to,
\begin{align}
    t_{\rm d} (\vec{x},\vec{y}) &= \frac{4GM_{\rm L}(1+z_{\rm L})}{c^3}\left[\frac{|\vec{x}-\vec{y}|^{2}}{2}-\psi(\vec{x}) \right]~,\label{eq:time_delay_EinsteinRadius}\\
    w &= \frac{8\pi G M_{\rm L}}{c^3}(1+z_{\rm L})\, f ~,\label{eq:dimensionless_frequency_EinsteinRadius}
\end{align}
respectively.

According to Fermat's principle, in the geometric optics limit ($\lambda_{\rm GW}\ll GM_{\rm L}/c^2$, or, $w\gg1$), the primary contribution to Eq. \eqref{eq:ampfac_dimensionless_fd} comes from the stationary points of $\tau_{\rm d}$. These stationary points are the location of images and can be determined using: \begin{equation}\label{eq:lens_equation}
\vec{\nabla}_{x}\tau_{\rm d}(\vec{x}, \vec{y})=0.
\end{equation}
For axially symmetric lenses, that is, $\psi(\vec{x}) = \psi(x)$, it can be seen that the images are collinear with $\vec{y}$. Therefore, without loss of generality, we choose $\vec{y}=(y, 0)$. In this study, we focus on the lens parameter space that generates multiple images in the geometric optics regime. The lens models examined here can produce at most two images. Cases with a single image are excluded, as their corresponding $\Ft$ has a straightforward functional form, making them trivial to model. We emphasize that the geometric optics quantities, such as, image time delays and image magnifications, are used in Sections~\ref{sec:peak_reconstruction} and \ref{sec:time_domain_ampl_transf} solely to improve the modeling of $\Ft$. This is because the primary contribution to $\Ff$ comes from the images, making it essential to capture them accurately. However, we do not apply the geometric optics approximation to $\Ff$ anywhere in this study.\footnote{In the geometric optics approximation, the parity of the images introduces a \textit{Morse phase} in the amplification factor. Since we do not impose such an approximation, this phase term is implicitly present in $\Ff$.}

\begin{figure*}[t]
    \centering
    \includegraphics[scale=1]{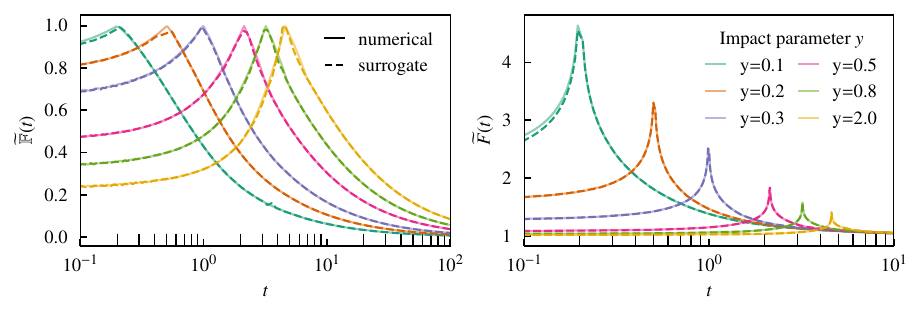}
    \includegraphics[scale=1]{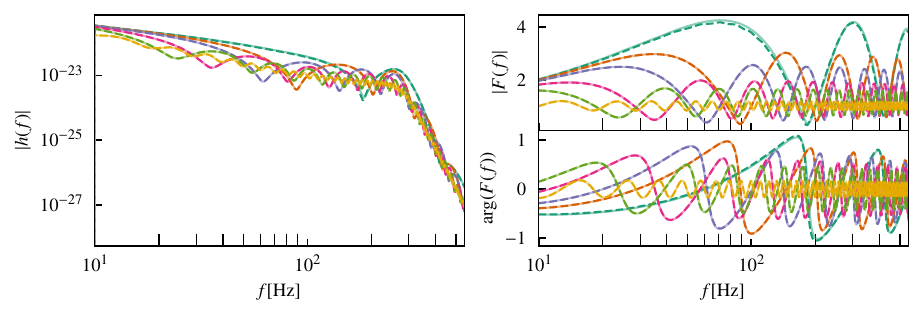}
    \caption{\textit{Top row}: The time-domain amplification factor, $\widetilde{F}(t)$, and its regularized version defined in Eq.~\eqref{eq:Ft_transf}, $\widetilde{\mathbb{F}}(t)$, due to a point-mass lens for various impact parameters $y$ as a function of time $t$ (in units of $4GM_{\rm L}(1+z_{\rm L})/c^3$, where the global minima lies at $t=0$). \textit{Bottom row}: Amplitude of the lensed waveforms (left) considering a GW150914-like source and the amplitude and phase of the frequency-domain amplification factor $F(f)$ (right) for corresponding values of impact parameters computed numerically (solid lines) and with the surrogate models (dashed lines) for a lens of mass $M_{\rm L}=10^3 M_{\odot}$ at redshift $z_{\rm L}=0.05$.}
    \label{fig:ampfac_PML}
\end{figure*}

\begin{figure}[h]
      \centering
      \includegraphics[scale=1]{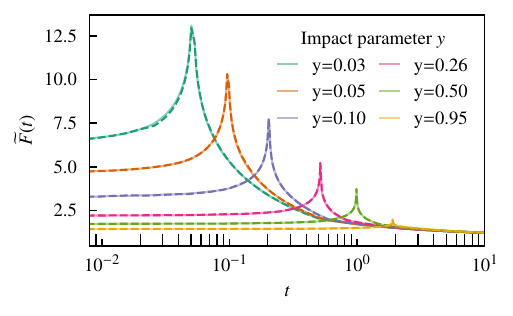}
      \caption{$\widetilde{F}(t)$ for various impact parameters $y$ for the SIS lens model computed using the numerical method (solid lines) and the surrogate model (dashed lines) as a function of time $t$ (in units of $4GM_{\rm L}(1+z_{\rm L})/c^3$, where the global minima lies at $t=0$).}
      \label{fig:training_Ft_SIS}
\end{figure}

In this work, we construct surrogate lensing waveforms for the point mass lens (PML) and the singular isothermal sphere (SIS) models, which are described in Appendix~\ref{appx-a}. Fig. \ref{fig:ampfac_PML} shows the amplification factor in both the time domain (top right panel) and frequency domain (bottom right panel) for various impact parameters $y$, computed numerically and using the surrogate model for PML (see Section \ref{sec:results} for details). The corresponding lensed GW waveforms are shown in the bottom left panel. The logarithmic peaks in $\widetilde{F}(t)$ (top right panel) occur at $t=t_{\rm peak}$ (see Appendix~\ref{appx-a}), a feature characteristic of the saddle image. Similarly, Fig. \ref{fig:training_Ft_SIS} illustrates the time-domain amplification factor, $\Ft$, computed for various impact parameters using both the numerical method and the surrogate model for SIS.

\subsection{Peak reconstruction}\label{sec:peak_reconstruction}
\begin{figure}[t]
      \centering
      \includegraphics[scale=1]{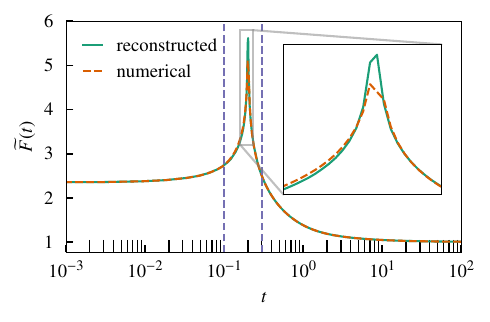}
      \includegraphics[scale=1]{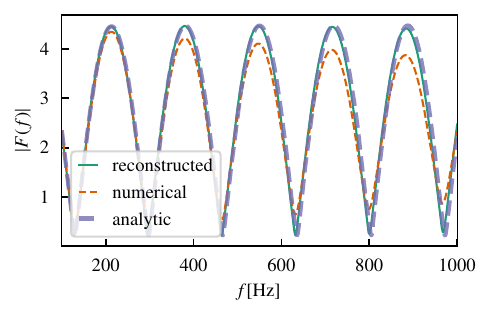}
      \caption{Illustration of improvement in the accuracy of $F(f)$ with peak reconstruction. \textit{Top}: Time domain amplification factor with and without peak reconstruction for PML lens for $y=0.1$ as a function of time $t$ (in units of $4GM_{\rm L}(1+z_{\rm L})/c^3$, where the global minima lies at $t=0$). The dashed vertical lines denote the window in which peak correction is applied. \textit{Bottom}: Amplitude of the frequency domain amplification factor with and without peak reconstruction for the same lens with mass $M_{\rm L}=1000 M_\odot$ at $z_{\rm L}=0.5$. Additionally, the analytically computed $|F(f)|$ is shown for comparison.
      }
      \label{fig:peak_correction_PML}
\end{figure}

Accurate computation of the time-domain amplification factor via
Eq.~(\ref{eq:ampfac_td}) demands sufficiently high resolution in the numerical
integration~\footnote{$F(w)$ is highly oscillatory for any general lens model, making it difficult to numerically compute using Eq.\eqref{eq:ampfac_dimensionless_fd}, as well as to model using surrogates. A better strategy is to evaluate $\Ft$ and then arrive at $F(w)$ or $F(f)$. This is also useful in generalizing the procedure to more complicated lenses. The discussion below Eq.\eqref{eq:time_delay_dimensionless} elaborates this point.}. Nonetheless, even at high resolution, the numerical solution may
struggle to capture the peak of $\widetilde{F}(t)$ due to the logarithmic
divergence at the peak. This limitation can impact the accuracy (especially at large $w$) of the frequency-domain amplification factor $F(w)$, as defined in
Eq.~(\ref{eq:FwFFT}) (see Fig. \ref{fig:peak_correction_PML} for an illustration). Since $w$ increases with $M_{\rm L}$ (refer to Eq.~\eqref{eq:dimensionless_frequency_EinsteinRadius}), accurate modeling of the peak becomes even more important for large $M_{\rm L}$ values.

To overcome the limitation of numerical methods in resolving the peak feature,
we utilize the analytical expressions for the peak location and the behavior of
$\widetilde{F}(t)$ in the vicinity of the peak given by \cite{ulmer1995},
\begin{equation}\label{eq:Ft_near_peak}
    F_{\rm approx}(t)=\sqrt{\mu_{+}}-\frac{1}{\pi}\sqrt{|\mu_{-}|}\;\ln\left(|t - t_{\rm peak}|\right) ~,
\end{equation}
where the values of $\mu_{+}, \mu_{-}$ and $t_{\rm peak}$ are given by Eq. \eqref{eq:img_mag_PML}, \eqref{eq:time_delay_PML} for PML and \eqref{eq:img_mag_SIS}, \eqref{eq:time_delay_SIS} for SIS lenses. We use the following
steps to reconstruct the peak features by smoothly hybridizing the numerical
and analytical approximation on either side of the peak:

\begin{enumerate}[label=(\alph*)]
\item Identify the peak location, $t_{\text{peak}}$, using the analytical
expression given in Eq. \eqref{eq:time_delay_PML} for PML and Eq. \eqref{eq:time_delay_SIS} for SIS.

\item Split the numerical $\widetilde{F}(t)$ at $t_{\text{peak}}$ into a left
segment, $F_{\text{num}}^{\text{l}}(t)$, and a right segment,
$F_{\text{num}}^{\text{r}}(t)$.

\item Using the analytical approximation of $\widetilde{F}(t)$ near the peak 
(Eq. \eqref{eq:Ft_near_peak}) evaluate the left and right
approximations, $F_{\text{approx}}^{\text{l}}(t)$ for $t < t_{\text{peak}}$ and
$F_{\text{approx}}^{\text{r}}(t)$ for $t > t_{\text{peak}}$.

\item Define a reconstruction window, $[t_{\text{match}}^{\text{l}},
t_{\text{match}}^{\text{r}}]$, around $t_{\text{peak}}$ to match and
reconstruct the peak. The window span is empirically chosen: a larger span reduces accuracy of the approximation in Eq.\eqref{eq:Ft_near_peak} as it moves away from the peak, while a smaller window span depends on the accuracy of the numerical method near the saddle image. A window spanning a time interval equal to $5\%$ on each side of $t_{\text{peak}}$ is found to be effective.

\item Rescale the approximated segments $F_{\text{approx}}^{\text{l}}(t)$ and
$F_{\text{approx}}^{\text{r}}(t)$ to match the numerical $\widetilde{F}(t)$
values at $t_{\text{match}}^{\text{l}}$ and $t_{\text{match}}^{\text{r}}$,
respectively.

\item Replace the portion of the numerical $\widetilde{F}(t)$ within the
reconstruction window with the rescaled approximated segments. 
%
\end{enumerate}

Following the above procedure results in a reconstructed time-domain amplification factor,
\[
    \widetilde{F}(t) = 
    \begin{cases}
        F_{\text{num}}^{\text{l}}(t), & t \leq t_{\text{match}}^{\text{l}} \\
        F_{\text{approx}}^{\text{l}}(t), & t_{\text{match}}^{\text{l}} < t < t_{\text{peak}} \\
        F_{\text{approx}}^{\text{r}}(t), & t_{\text{peak}} < t < t_{\text{match}}^{\text{r}} \\
        F_{\text{num}}^{\text{r}}(t), & t \geq t_{\text{match}}^{\text{r}}
    \end{cases} \,,
\]
which is a continuous, but non-smooth, function. Fig. \ref{fig:peak_correction_PML} shows that reconstructing the peak in the $\Ft$ (top panel) consistently provides a better match (bottom panel) with the analytic expression (Eq. \eqref{eq:ampfac_PML_analytic}) as compared to the vanilla numerical $\widetilde{F}(t)$. As seen in the figure, the difference between the various methods becomes more pronounced at higher frequencies, primarily due to limited resolution near the peak. However, as discussed in Section~\ref{sec:results}, the peak-reconstructed method has been calibrated to ensure a maximum mismatch of $10^{-4}$ within the parameter space. Increasing the resolution can improve accuracy at high frequencies, though this comes at the cost of longer generation times for the training data. Importantly, this does not impact the evaluation speed of the surrogate model.

\subsection{Time-domain amplitude regularization}\label{sec:time_domain_ampl_transf}

The logarithmic singularity at $t_{\rm peak}$ presents a challenge for accurately modeling the region near the peak of $\Ft$ using a surrogate. To address this, we apply a regularization procedure to the reconstructed $\Ft$ (see Section~\ref{sec:peak_reconstruction} for details on the reconstruction) and construct a surrogate model based on this peak-reconstructed, regularized time-domain amplification factor, $\Ftt$. The following transformation suppresses the peak contribution:
\begin{equation}\label{eq:Ft_transf}
    \widetilde{\mathbb{F}}(t) = 1 - \exp\left[-\pi\frac{\left(\Ft-1\right)}{\sqrt{\mu_{-}}}\right] ~,
\end{equation}
where $\mu_{-}$ is the magnification due to the saddle image, given by Eq. \eqref{eq:img_mag_PML} for PML and \eqref{eq:img_mag_SIS} for SIS lens models. The regularization effectively eliminates the singularity and makes the derivative of the amplification factor continuous. The top left panel in Fig. \ref{fig:ampfac_PML} compares the regularized $\Ftt$ for PML computed for various impact parameters using the numerical method and the surrogate model for PML. After evaluating the surrogate, an inverse transformation is applied to obtain the time domain amplification factor $\Ft$ from $\Ftt$.
\subsection{Surrogate Modeling}
\label{sec:surrogate_modeling}

In this section, we introduce the basics of surrogate modeling, its application
in computing the lensing amplification factor for lensed GW waveforms and the
setup used in this work. Surrogate modeling has been used extensively in
building fast and accurate GW waveforms, especially in building surrogate
models for Numerical Relativity (NR)
waveforms~\cite{Field:2013cfa,Blackman:2015pia,Tiglio:2021ysj}. Surrogate modeling is particularly advantageous in situations where
the number of available waveforms is limited due to the prohibitively
large computation time required to generate them, such as
when solving partial differential equations, which results
in a sparse training dataset. It is also highly beneficial when
a data analysis study demands millions of waveform evaluations, 
requiring exceptionally rapid model evaluation times to meet the
computational demands efficiently. For example, effective-one-body 
surrogates~\cite{Purrer:2014fza,Purrer:2015tud,Cotesta:2020qhw,Gadre:2022sed,Khan:2020fso,Thomas:2022rmc} are extensively used as part of LVK parameter estimation efforts.

In the following, we will briefly discuss how surrogate modeling reduces the
dimensionality of the problem by using reduced basis and empirical
interpolation methods. These two methods
ensure that with only a limited number of waveforms (chosen appropriately), one
can build an accurate model for predicting solutions outside the training set
of parameters. Because these two steps can be precomputed, it also ensures that
the model, evaluated at arbitrary points in parameter space, is orders of
magnitude faster than numerical solutions.

\subsubsection{Surrogate modeling basics}
\label{sec:surrogate_modeling_basics}
We follow the methods described in Sec.~II and III of Ref.~\cite{Field:2013cfa}
and refer the reader to it for a detailed overview of building a surrogate
model. In the following, we briefly outline building a surrogate
model.

\begin{enumerate}
\item The first step is to find a minimal set of solutions called the {\itshape
reduced basis} (RB)~\cite{Field:2011mf} in terms of which solutions at other
parameters can be expressed. Given a known set of solutions (also called the
training set), the RB is found using a greedy search algorithm. For example,
with a RB with $m$ basis elements, $\{e_{i}\}_{i=1}^{m}$, a solution $\widetilde{\mathbb{F}}(t;\boldsymbol{\lambda})$ in the
training set at the lens parameter values $\boldsymbol{\lambda}$ can be well
approximated as:
  \begin{equation}
    \label{eq:expansion_in_rb}
    \widetilde{\mathbb{F}}(t; \boldsymbol{\lambda}) \approx \sum_{i=1}^{m} c_i(\boldsymbol{\lambda}) \,e_{i}(t).
  \end{equation}
  This approximation is also good at other parameters outside the training set
as long as the training set is dense enough. However, getting a dense training
set can be very expensive and sometimes prohibitive. In such cases, one
can build the training set using a greedy search method to provide an
optimal training set for a given desired accuracy of the surrogate
prediction. We come back to this in sec.~\ref{sec:surrogate_modeling_setup}
where we discuss the surrogate modeling setup specific to our problem.

Since the number of RB, $m$, is usually very small compared to the number of solutions in the training set, say $M$, the dimensionality of the problem reduces by a factor of $m/M$, which is usually $\ll 1$.

\item The next step is to find the most representative times, the empirical
times or nodes $\{T_i\}_{i=1}^{m}$ to construct an interpolant in time using
{\itshape empirical interpolant method} (EIM)~\cite{barrault2004empirical,YvonMayday,Hesthaven2014,Field:2013cfa,chaturantabut2009discrete} for
a given parameter $\boldsymbol{\lambda}$:
\begin{equation}
  \label{eq:EIM}
  \mathcal{I}_{m}[\widetilde{\mathbb{F}}](t; \boldsymbol{\lambda}) = \sum_{i=i}^{m} C_{i}(\boldsymbol{\lambda})\, e_{i}(t).
\end{equation}
The coefficients $C_{i}(\boldsymbol{\lambda})$ are defined by requiring that
the interpolant becomes equal to the value of the solutions at the empirical
nodes:
\begin{equation}
  \label{eq:EIM-coefficients}
  \sum_{i=i}^{m} C_{i}(\boldsymbol{\lambda})\,e_{i}(T_j) = \widetilde{\mathbb{F}}(T_j;\boldsymbol{\lambda}),\qquad j=1,...,m
\end{equation}
or equivalently,
\begin{equation}
  \label{eq:EIM-Vij}
  \sum_{i=1}^{m} V_{ji}\, C_{i}(\boldsymbol{\lambda}) = \widetilde{\mathbb{F}}(T_j; \boldsymbol{\lambda}),\qquad j=1,...,m
\end{equation}
where the interpolation matrix is given by
\begin{equation}
  \label{eq:V}
  V \equiv
  \begin{pmatrix}
    e_{1}(T_1) & e_{2}(T_1) & \ldots & e_m(T_1) \\
    e_{1}(T_2) & e_{2}(T_2) & \ldots & e_m(T_2) \\
    \vdots & \vdots & \ddots & \vdots \\
    e_{1}(T_m) & e_{2}(T_m) & \ldots & e_m(T_m) \\
  \end{pmatrix}.
\end{equation}
The coefficients $C_{i}(\boldsymbol{\lambda})$ can be obtained by solving the above $m$-by-$m$ system: 
\begin{equation}
  \label{eq:C}
  C_{i} = \sum_{j=1}^{m} (V^{-1})_{ij} \,\widetilde{\mathbb{F}}(T_j;\boldsymbol{\lambda}).
\end{equation}
Substituting Eq.~(\ref{eq:C}) in Eq.~(\ref{eq:EIM}), the empirical interpolant can be rewritten as: 
\begin{equation}
  \label{eq:EIM-in-B}
    \mathcal{I}_{m}[\widetilde{\mathbb{F}}](t; \boldsymbol{\lambda}) = \sum_{j=i}^{m} B_{j}(t)\,\widetilde{\mathbb{F}}(T_j;\boldsymbol{\lambda}),
  \end{equation}
  with:
  \begin{equation}
    \label{eq:B}
    B_{j}(t) \equiv \sum_{i=1}^{m} e_{i}(t)\,(V^{-1})_{ij}.
  \end{equation}

The function $B_j(t)$ is independent of $\boldsymbol{\lambda}$ 
and can be precomputed offline using only the information
contained in the RB. Similar to the RB, the empirical nodes -- also 
independent of $\boldsymbol{\lambda}$ and equal in number to the RB 
functions -- are determined using a separate greedy search algorithm. 
Since the RB typically consists of only a few functions, 
the empirical interpolant's representation of the function, as expressed 
in Eq.~\eqref{eq:EIM-in-B}, requires only a sparse set of time points.
This eliminates the need for a dense, uniform temporal grid, which would
otherwise significantly increase computational cost. As a result, the
temporal dimensionality of the problem is reduced by a factor of 
$m/L$, where $m$ is the number of RB functions (and empirical nodes), 
and $L$ is the number of points required for interpolation on a uniform grid.

The use of RB and EIM reduces the dimensionality of the problem by a factor of
$(m \times m)/(M \times L) \ll 1$.

\item With the interpolant $\mathcal{I}_{m}[\widetilde{\mathbb{F}}](t; \boldsymbol{\lambda})$ in
Eq.~(\ref{eq:EIM-in-B}) at hand, we need to know the values of the solutions at
the empirical nodes $\{T_{i}\}_{i=1}^{m}$. Therefore, the third step is
to construct parametric fits at each empirical node across the parameter space.

\item Finally, the surrogate model prediction at a new parameter
$\boldsymbol{\lambda_{\ast}}$ outside the training data set is computed by
first using the parametric fits (constructed in the third step) to evaluate the
values of $\widetilde{\mathbb{F}}(t; \boldsymbol{\lambda_{\ast}})$ at the empirical nodes
$\{T_{i}\}_{i=1}^{m}$ and then building the interpolant $\mathcal{I}_{m}[\widetilde{\mathbb{F}}](t; \boldsymbol{\lambda})$ using Eq.~(\ref{eq:EIM-in-B}).
\end{enumerate}

The first three steps are performed offline (i.e, precomputed), which drastically expedites the evaluation of the surrogate model at arbitrary points in parameter space. 

\subsubsection{Surrogate for microlensed waveforms}
\label{sec:surrogate_for_microlensed_waveforms}
We apply the technique of surrogate modeling
to obtain a model of the GWs waveform undergoing microlensing. However, since
the effect of microlensing is a frequency-dependent modulation of the GW
amplitude as explained in Sec.~\ref{sec:lensing_amplification_factor}, one
needs only to model this amplification factor provided a fast, unlensed model
exists already. Therefore, instead of modeling the microlensed waveform
directly, we build a surrogate model $\sFtt$ for the regularized lens amplification factor
$\Ftt$ in the time domain. We then obtain the frequency-dependent amplification
factor $\sFf$ by taking a Fourier transform of the back-transformed $\sFt$ (see Sec. \ref{sec:time_domain_ampl_transf}). The surrogate
microlensed waveform $\sLhf$ is then obtained by multiplying the unlensed
waveform $\hf$ by the frequency-dependent amplification factor $\sFf$:
\begin{equation}
  \label{eq:lensed_surrogate_waveform}
  \sLhf = \sFf \times \hf.
\end{equation}

\subsubsection{Surrogate modeling setup}
\label{sec:surrogate_modeling_setup}
The RB is found by using a greedy algorithm that guarantees that the error
\begin{align}
  \label{eq:sigma_m}
  \sigma_m \equiv \max_{\lambda}\min_{c_i\in C}\left\lvert\left\lvert\widetilde{\mathbb{F}}(.; \boldsymbol{\lambda}) - \sum_{i=1}^{m} c_i(\boldsymbol{\lambda}) \,e_{i}(.)\right\rvert\right\rvert \,,
\end{align}
associated with the approximation in Eq.~(\ref{eq:expansion_in_rb}) (minimized
over the coefficients $c_i$ and maximized over the parameter
$\boldsymbol{\lambda}$) is less than a desired value, usually called the basis
tolerance. Here, we have defined the standard $L_2$ norm as
\begin{align}
  \quad \left\lvert\left\lvert \widetilde{\mathbb{F}}(.; \boldsymbol{\lambda})  \right\rvert\right\rvert \equiv \sqrt{\int_{t_{\text{min}}}^{t_{\text{max}}} \left\lvert\widetilde{\mathbb{F}}(t; \boldsymbol{\lambda})\right\rvert^2 \dd t}\,.
\end{align}
Using an optimal basis tolerance when building the RB is crucial in
making a surrogate model as fast as possible. Using a very small basis
tolerance may result in a larger number of bases, which in turn results in a
larger number of empirical time nodes. While evaluating the surrogate model, a
significant portion of the computation time involves evaluating the fits at the
empirical nodes. Thus, a smaller basis tolerance is likely to make the
surrogate model slower. Therefore, we use an optimal basis tolerance of
$\roughly 5\times 10^{-4}$ for building the RB so that the surrogate models achieve the
required accuracy compared to the microlensed waveform computed using the
numerical solutions without causing any overfitting. 

To obtain the training data set, we employ a greedy search algorithm
as described in \cite{Blackman_2017}. In the
greedy search, we start with an initial training set consisting of only the
corner points of the parameter range of interest. At each step, we build the
surrogate model with a specific basis tolerance ($\roughly 5\times 10^{-4}$ in our
case) and then validate it against a large number of numerical solutions
generated randomly within the parameter range. This validation set excludes the
parameters already in the training set. The validation involves measuring the
normalised $L_2$ error between the surrogate prediction $\sFtt$ and the
numerical solution $\Ftt$,
\begin{equation}
  \label{eq:L2}
  L_2 = \frac{\sqrt{\int_{t_{\text{min}}}^{t_{\text{max}}} \left|\sFtt - \Ftt\right|^2 \dd t}}{\sqrt{\int_{t_{\text{min}}}^{t_{\text{max}}} \left|\Ftt\right|^2 \dd t}},
\end{equation}
where $t_{\text{min}}$ and $t_{\text{max}}$ denote the range of the time segment within which the error is computed.

We pick the parameter where the $L_2$ error is the maximum and add it to the
training set to be used in the next step and repeat the procedure until we
achieve the maximum $L_2$ error to be smaller than a desired value, which in our
case is $10^{-2}$. With such a greedy search, we find our training data sets that
consist of $\roughly 35-100$ data points depending on the particular microlens model.

For constructing the fits at the empirical nodes, we use the Gaussian Process
Regression (GPR) method as described in the supplemental material of \cite{Varma:2018aht}. 

In a nutshell, the training data $\Ftt$ is obtained by applying a peak reconstruction followed by a regularization operation on $\Ft$. This training data is used to build the surrogate $\sFtt$. Then an inverse transformation is applied followed by a peak reconstruction to obtain $\sFt$.


\section{Results}
\label{sec:results}
In order to validate our model, we compute the \emph{match} $\mathcal{M}$
between the surrogate microlensed waveform model and the corresponding numerical model used for training. Match between two given waveforms $\h_1$
and $\h_2$ is defined as \cite{LIGOScientific:2019hgc}:
\begin{equation}
  \label{eq:match}
  \mathcal{M} \equiv \arg \max_{t_c, \phi_c}\frac{\langle\h_1, \h_2\rangle}{\sqrt{\langle\h_1, \h_1\rangle\langle\h_2, \h_2\rangle}},
\end{equation}
where $t_c$ and $\phi_c$ are the coalescence time and phase, respectively, and
the overlap $\langle \h_1, \h_2\rangle$ is:
\begin{equation}
  \label{eq:overlap}
  \langle \h_1, \h_2\rangle = 4 \mathcal{R} \int_{f_{\text{low}}}^{f_{\text{high}}} \frac{\tilde{\h_1}^\star(f) \tilde{\h}_2(f)}{S_n(f)}\text{d} f.
\end{equation}
Here, $f$ is the frequency, $\tilde{\h}_{1,2}$ are the Fourier transforms of
the corresponding time domain waveforms, $\star$ represents the complex
conjugate and $\mathcal{R}$ represents the real part. $f_{\text{low}}$ and
$f_{\text{high}}$ denote the lower and higher frequency cutoff of the detector
band and $S_{n}(f)$ denote the one-sided power spectral density (PSD) of the
detector noise. We use \texttt{pycbc.filter.matchedfilter} module from the \texttt{PyCBC} package \cite{pycbc} to compute the match. We make use of the next-generation ground-based detector Einstein Telescope (ET) PSD, pertaining to the ET design sensitivity \cite{Hild:2010id}. The reason for choosing this particular PSD is twofold: the frequency bin is larger than the current ground-based detectors, giving rise to conservative mismatches, and statistically, we expect to see a significant number of lensed GW events during the observing runs of next-generation detectors such as ET.

Note that, the numerical amplification factors used to train the model for PML and SIS are calibrated to the analytic form given in Eq. \eqref{eq:ampfac_PML_analytic} for PML and the summation form given in Eq. \eqref{eq:ampfac_SIS_summation} for SIS, such that the maximum mismatch ($1 - \mathcal{M}$) between the numerically computed lensed waveform and the analytically obtained lensed waveform is $<10^{-4}$ for all possible points in the parameter space.

We present mismatches between the surrogate microlensed GWs and the numerically evaluated microlensed GWs used to construct the surrogate model. We ensure that the lensing parameters used for training the model, and those used for testing (i.e, evaluating the mismatches), are different. The lensing parameters are chosen over a grid, and the mismatches are evaluated for each point on that grid, as well as different unlensed CBC waveforms. Evaluation times per microlensed frequency-domain waveform are also measured. These are then compared with corresponding times pertaining to other methods, viz., a frequency domain interpolation method employed by the LVK collaboration to search for GWs microlensed by PMLs, as well as numerical methods to acquire amplification factor for the SIS lens. The waveforms are sampled at frequencies with bin width, $\delta f = 1/32$, resulting in the total number of waveform evaluation points in the frequency domain to range from $\roughly 1.3\times 10^4$ (for total binary mass $M_{\rm tot}=100M_\odot$) to $\roughly 6.5\times 10^4$ (for total binary mass $M_{\rm tot} = 20 M_\odot$). 

The surrogate model for PML is created using 27 basis vectors with a basis tolerance of $5\times 10^{-4}$, while for SIS, it is created using 15 basis vectors with the same basis tolerance. We use the \texttt{numpy.fft} module from \texttt{NumPy} \cite{Harris:2020xlr} package to compute FFT. All computations related to creating and evaluating the surrogate model were performed on a 32-core workstation with Intel Xeon E5-2650 v2 chip, OS: Linux. Note that no parallelization was used during the evaluation; thus, the computations effectively used a single core. 

\begin{figure}[t]
      \centering
      \includegraphics[scale=1]{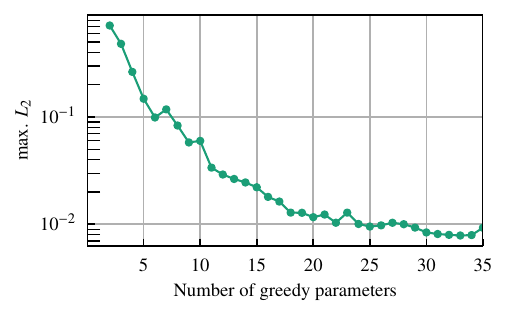}
      \includegraphics[scale=1]{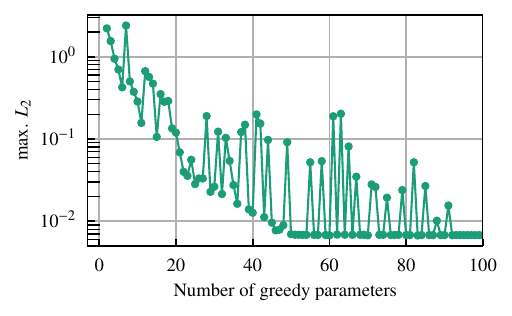}
      \caption{ Largest relative $L_2$ error of the surrogate for the PML model (top) and the SIS model (bottom) as a function of number of greedy parameters. The error is computed as the maximum error between the entire validation set and each surrogate model. Both surrogate models are seen to converge as the training data set increases.}
      \label{fig:greedy_mismatch_PML}
\end{figure}

\paragraph{Point Mass Lens (PML):} We create a surrogate model for the regularized time-domain amplification factor $\Ftt$. It is a single-parameter model parameterized by the impact parameter $y$. Note that the lens mass $M_{\rm L}$ sets the time scale while converting the dimensionless angular frequency $w$ to the dimension-full frequency $f$. Therefore, it can be factored out while creating the model. We use the numerically computed $\Ftt$ (as shown in Fig. \ref{fig:ampfac_PML}) to train our model. Fig. \ref{fig:greedy_mismatch_PML} shows the improvement in the maximum $L_2$ between the numerically obtained $\Ftt$ and the surrogate $\sFtt$ with an increase in the number of greedy parameters. The error versus greedy parameters curve plateaus for greedy parameter counts greater than approximately 30 for PML. Therefore, we use the first 35 greedy parameters to construct the surrogate model. The greedy parameter space of $y$ spans from $y=0.1$ to $y=2$.

\begin{figure}[h]
      \centering
      \includegraphics[scale=1]{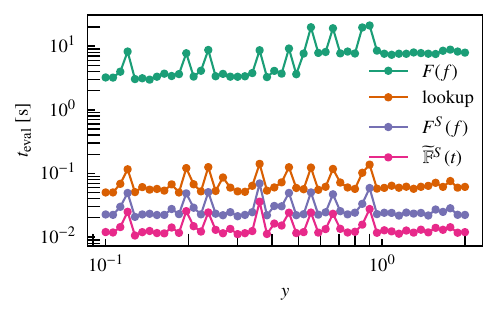}
      \caption{Evaluation time for the amplification factors due to point mass lens of mass $M_{\rm L}=100 M_\odot$ at $z_{\rm L}=0.5$, for various impact parameters $y$. The source is an equal mass CBC source with $M_{\rm tot}=20M_\odot$. We compare the evaluation times for the numerical amplification factor $\left(\Ff\right)$, surrogate model in the time domain $\left(\sFtt\right)$ and after FFT $\left(\sFf\right)$ with the time taken using the lookup table.}
      \label{fig:time_comparison_PML}
\end{figure}

The current LVK microlensing analysis makes use of an interpolation table (``lookup table'') in $w-y$ plane to compute $F(w)$ given in Eq.~\eqref{eq:ampfac_PML_analytic} for PML and Eq.~\eqref{eq:ampfac_SIS_summation} for SIS (see, e.g., ~\cite{Wright:2021cbn}). We compare the evaluation time of the PML surrogate model with the lookup table to verify that it performs at least as well as the lookup table for simple lensing cases. For the SIS surrogate model, we compare with the numerical method, as our goal is to extend this study to more complex lens models where interpolation tables may not be feasible, and only direct numerical computations are possible.

Fig. \ref{fig:time_comparison_PML} shows the evaluation time for the surrogate model compared to the lookup table for the same number of evaluation points. We also indicate the time taken to generate the numerically obtained $\Ff$. As seen from the figure, the surrogate performs faster than the lookup table throughout the parameter space. Note that a part of the lookup table evaluation uses information from the geometric optics approximation, whereas the numerical method and therefore, the surrogate model is based completely on the wave optics computation.

\begin{figure}[h]
      \centering
      \includegraphics[scale=1]{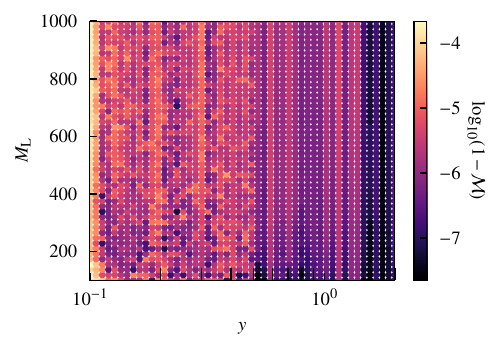}
      \caption{Mismatch in the $M_{\rm L} - y$ plane between the numerical method and the surrogate model for a lensed GW150914-like signal. We assume the lens to be a point mass lens at $z_{\rm L}=0.05$}.
      \label{fig:mismatch_GW150914}
\end{figure}

Fig. \ref{fig:mismatch_GW150914} shows the mismatch between the surrogate microlensed waveform and the microlensed waveform obtained using the numerical method for a GW150914-like signal for various values of $M_{\rm L}$ and $y$. The unlensed waveform is generated using the \texttt{IMRPhenomXP} approximant \cite{Pratten:2020ceb}. As evident from the figure, the mismatches are better than $0.03\%$.

\begin{figure*}[t]
    \centering
    \includegraphics[scale=1]{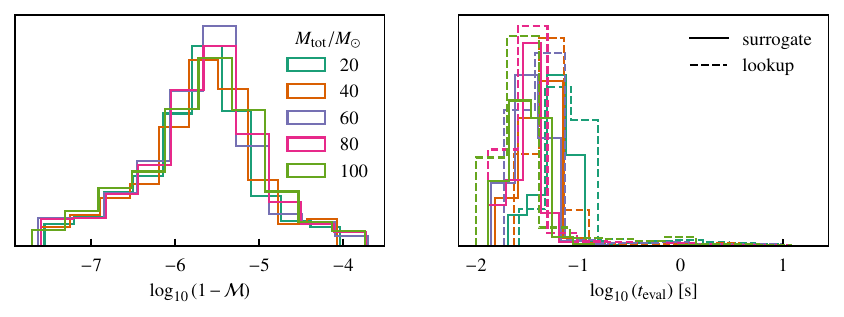}
    \includegraphics[scale=1]{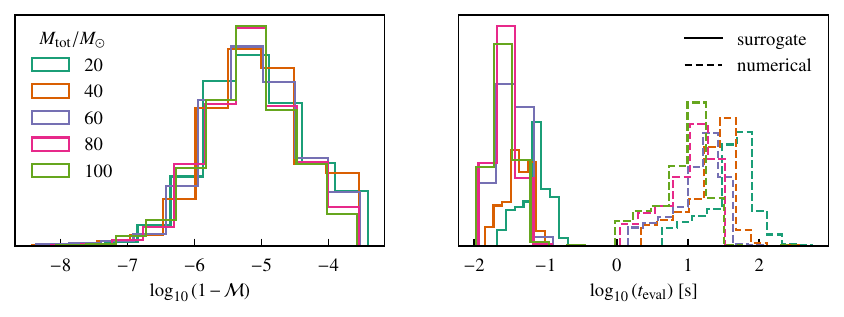}
    \caption{\textit{Top panel}: (\textit{Left}) Histograms of mismatch between the lensed waveforms due to PML computed using the surrogate model and the numerical method. Each histogram corresponds to a different source with the total source mass, $M_{\rm tot} = \{20, 40, 60, 80, 100\} M_\odot$. The mismatches are computed in a grid of $M_{\rm L}-y$, where $100 \leq M_{\rm L}/M_\odot \leq 1000$ and $0.1\leq y\leq 2$. (\textit{Right}) Histograms of the evaluation time for the surrogate model (solid) and for the lookup table (dashed) in the same $M_{\rm L}-y$ grid for PML. \textit{Bottom panel}: (\textit{Left}) Similar to the top left plot but for SIS lens with the range of $y$ being $0.01\leq y\leq 0.95$. (\textit{Right}) Histograms of the evaluation time for the surrogate model (solid) and for the numerical method (dashed) in the same $M_{\rm L}-y$ grid for SIS.}
    \label{fig:MM_comptime_all}
\end{figure*}

Further, we show the dependence of the accuracy of our model on the source mass of the binary by simulating equal component mass, non-spinning, binary black hole (BBH) unlensed waveforms for different total binary masses ($M_{\rm tot} = \{20, 40, 60, 80, 100\}M_\odot$) by computing mismatch in the $M_{\rm L} - y$ grid spanning $100\leq M_{\rm L}/M_\odot\leq 1000$ and $0.1\leq y\leq 2$. We consider equal mass binaries. The top left panel in Fig. \ref{fig:MM_comptime_all} shows the histogram of mismatch values for the different source masses between the lensed waveforms generated using the numerical method and the surrogate model. As can be seen, the mismatches are lower than $0.05\%$. Moreover, the top right panel compares the evaluation times for the surrogate with the lookup table for the same grid and source masses. 

\paragraph{Singular Isothermal Sphere (SIS):}\label{sec:results_SIS}
Similar to the PML model, this model is also a single-parameter model parameterized by $y$. As shown in Fig. \ref{fig:greedy_mismatch_PML}, the error versus greedy parameters curve plateaus for greedy parameter counts greater than approximately 85 for SIS. Therefore, to construct the surrogate model for SIS, we use the first 100 greedy parameters. The greedy parameter space of $y$ spans from $y=0.01$ to $y=0.95$. We compute mismatch and compare evaluation times between the numerically obtained microlensed waveforms and the surrogate microlensed waveforms. The bottom left panel in Fig. \ref{fig:MM_comptime_all} shows the histogram of mismatch values for various source masses, similar to the case for PML. Here, we restrict the $y$ values between $0.01\leq y\leq 0.95$ because for $y> 1$ we are no longer in the two-image regime. Similar to PML, the mismatch values are within $0.05\%$, showing very promising results. The bottom right panel compares the evaluation times for the surrogate with the numerically obtained microlensed waveform. The surrogate model is evidently faster than the numerically obtained microlensed waveforms by orders of magnitude. 


\section{Summary and Discussion}
\label{sec:conclusions}

Gravitational microlensing of GWs is likely to be detected in the near future — if not in O4, then in upcoming observing runs of the LVK detector network or next-generation ground-based detectors. However, detection rates with current ground-based detectors remain highly uncertain \cite{lai2018, christian2018}. Such detections promise to shed light on a host of questions pertaining to astrophysics, cosmology, and fundamental physics. Beyond probing the IMBH population, diffraction effects play a crucial role in modeling signals originating near the critical curve(s) of a lens, such as a galaxy modelled by an SIS profile. Moreover, the high time resolution of ground-based detectors enables the exploration of lensing effects at scales beyond the reach of optical or radio telescopes, offering a unique avenue for discovering previously unobserved lens populations.

Probing for signatures of microlensing-related wave-optics effects in detected GW events requires large scale Bayesian PE runs, whose feasibility crucially relies on accurate and rapidly producible microlensed waveforms. Apart from GWs microlensed by a PML and a few other simplistic lensing models, no closed form analytical solutions to the lensing equation exist. Numerical methods are, therefore, the only recourse to generate microlensed waveforms for most realistic lensing configurations.

Evaluating microlensed GWs numerically is time consuming, with waveform generation taking up to several seconds (per waveform). Indeed, even efficient numerical methods (see, e.g: \cite{tambalo2023}) are likely to be too slow to enable large-scale PE campaigns in a cost-effective manner. One way to mitigate this issue is to use interpolation, where the microlensing amplification factor is evaluated at a few discrete points in the parameter space of the lens, and then interpolated to any arbitrary point. This has been attempted for the PML in the frequency domain, and found to be faster than directly evaluating the complicated amplification factor analytically \cite{LIGOScientific:2023bwz}. It has also been attempted for other spherically symmetric lensing configurations in the time domain, where the amplification factor is then Fourier transformed to the frequency domain \cite{Cheung:2024ugg}. Both these methods require reading-off precomputed interpolation data from a table, and are referred to in this work as the ``lookup table'' methods. For other efficient methods for rapid evaluation of lensing amplification factors, we refer the reader to \cite{Villarrubia-Rojo:2024xcj}.

Given the impressive success of surrogate modelling to accurately and rapidly produce (unlensed) CBC templates \cite{Field:2013cfa,Blackman:2017pcm,Varma:2019csw,Islam:2021mha,Yoo:2022erv,Yoo:2023spi,Gerosa:2018qay,Varma:2018aht,Islam:2023mob}, in this work, we assess its applicability in the context of generating microlensed GWs. We construct surrogate models of GWs microlensed by a PML and SIS lens. We consider a range of realistic lens masses and impact parameters. We achieve this by building time-domain surrogate amplification factors, Fourier transforming them, and then multiplying them with the unlensed frequency-domain GW waveform.

To assess the performance of the surrogate models, we compute mismatches between the surrogate microlensed waveforms and the corresponding, numerically evaluated, waveforms that were used to build the surrogate models, ensuring that the mismatches are not evaluated at lens parameter values used for surrogate construction. We do so over a grid of lens parameter values, and find that we're able to achieve mismatches in the range $\sim 10^{-8} - 10^{-4}$. Moreover, we measure the evaluation times of the microlensed waveforms in the frequency domain and find that they're typically $\mathcal{O}(10^{-2} - 10^{-1})$s for number of sample points ranging from $\roughly (1.3 - 6.5) \times 10^4$. These benchmarking tests demonstrate the power of surrogate modelling applied to microlensed GWs, suggesting that large-scale PE runs can be feasibly conducted using surrogate amplification factors.

It is worth pointing out that the rate-determining step in the computation of surrogate microlensed GWs in the frequency domain is, in fact, the Fourier transform, for which we employ \texttt{NumPy}'s FFT package \cite{Harris:2020xlr}. The evaluation times of the time-domain surrogate waveforms are smaller than the FFT compuation by a couple of orders of magnitude.  More sophisticated FFT algorithms may further reduce the evaluation times. Nevertheless, we still find that, for the PML, the microlensed waveforms computed using the lookup table method in the frequency domain is still slower than the corresponding surrogate waveforms by about an order of magnitude for large portions of the lensing parameter space considered.

The main advantage of surrogate modelling of microlensed GWs is that it does not intrinsically rely on any symmetries of the lensing potential. Indeed, if numerical solutions are able to provide time-domain amplification factors, it should be possible to construct corresponding surrogate models. Thus, in principle, our work should be readily extendable to asymmetric lensing configurations that better model those found in nature, such as a microlens embedded in a macropotential (e.g: an intermediate mass black hole inside a galaxy-scale lens). However, constructing surrogate models for realistic lenses needs to mitigate an important problem. 

Caustics -- curves in the source plane where the image magnification formally diverges -- demarcate distinct regions in that plane. These regions differ from each other in the number of images\footnote{in the geometric optics limit} that will be produced if the source is positioned in those regions. Abrupt changes in the number of images when crossing over caustics correspond to changes in the shape of the time-domain amplification factor as a function of the impact parameter that are not smooth. Indeed, even in the case of SIS lens, we restricted ourselves to lens parameters that ensure that exactly two images are produced. A possible workaround is to build separate surrogate models for each region in the lens plane containing a fixed number of images. However, an additional challenge pertaining to this workaround is to appropriately model the shape of the caustics. This should be straightforward for lensing potentials with symmetries, but might not be so for asymmetric lens systems, especially those that are an aggregation of multiple microlensing potentials embedded in a macropotential. We leave the surrogate modelling of such lensing configurations for future work. In addition, the accuracy of the model depends on the quality of the training data, which in turn relies on the numerical method and the image-finding algorithm (for peak reconstruction and amplitude regularization). Potential inaccuracies at high frequencies can be mitigated by increasing the resolution near the peaks of $\Ft$. Future work will focus on further refining these aspects for enhanced performance.

\begin{acknowledgements}
We are grateful to Tousif Islam for the careful review of the manuscript and useful comments. We thank the members of the astrophysical relativity group at the International Centre for Theoretical Sciences for their valuable input. We acknowledge the support of the Department of Atomic Energy, Government of India, under project no. RTI4001.
M.A.S.’s research was supported by the National Research 
Foundation of Korea under grant No. NRF-2021M3F7A1082056. 
S.J.K gratefully acknowledges support from SERB grant SRG/2023/000419. 
V.V.~acknowledges support from NSF Grant No. PHY-2309301.
S.E.F.~acknowledges support from NSF Grants PHY-2110496 and AST-2407454.
S.E.F.~and V.V.~were partially supported by UMass Dartmouth's Marine and Undersea
Technology (MUST) research program funded by the Office of Naval Research 
(ONR) under grant no. N00014-23-1-2141.
All computations were performed using the computational facilities at the International Centre for Theoretical Sciences and the Inter-University Centre for Astronomy and Astrophysics. 

\end{acknowledgements}

\appendix

\section{Lens models}\label{appx-a}
In this appendix, we describe the two lens models, namely, point mass lens (PML) and singular isothermal sphere (SIS) used in this work.

\subsubsection{Point Mass Lens}\label{sec:PML_amplification_factor}
For a point-mass lens (PML) with mass $M_{\rm L}$, the projected surface mass density on the lens plane is given by \cite{schneider2013gravitational}: 
\begin{equation}\label{eq:PML_sigma}
    \Sigma(\vec{x}) = \frac{M_{\rm L}}{\xi_0^2}\delta^2(\vec{x}) = \frac{M_{\rm L}}{\xi_0^2} \frac{1}{2\pi x}\delta(x).
\end{equation}
To determine the lensing potential, we solve the Poisson equation in two-dimensions:
\begin{equation}\label{eq:sigma_to_psi}
    \nabla^2_x\psi(\vec{x}) = \frac{2\Sigma(\vec{x})}{\Sigma_{\rm cr}},
\end{equation}
where:
\begin{equation}\label{eq:critical_density}
    \Sigma_{\rm cr} \equiv \frac{c^2}{4\pi G}\frac{D_{\rm S}}{D_{\rm L}D_{\rm LS}}.
\end{equation}
If we set $\xi_0$ to be the \emph{Einstein radius} of the lens,
\begin{equation}\label{eq:einstein_radius}
  \xi_0^2=4\frac{GM_{\rm L}}{c^2}\frac{D_{\rm L}D_{\rm{LS}}}{D_{\rm S}},
\end{equation}
the lensing potential for PML is found to be: 
\begin{equation}\label{eq:pml_lensing_potential}
    \psi_{\rm PML}(x) = \ln\left(x\right).
\end{equation}

As mentioned in the main text, the amplification factor for the PML can be analytically computed \cite{Peters:1974gj}:
\begin{align}
    F(w) = & \exp\left[\frac{\pi w}{4}+\frac{i w}{2}\left(\ln\left(\frac{w}{2}\right)-2\phi_m(y)\right)\right]\nonumber\\
    &\times \Gamma\left(1-\frac{i}{2}w\right)\hspace{2pt}_{1}{F}_1\left(\frac{i}{2}w,1;\frac{i}{2}wy^2\right),\label{eq:ampfac_PML_analytic}
\end{align}
where:
\begin{align*}
    \phi_m(y) &= \frac{(x_m-y)^2}{2}-\ln x_m,\\
    x_m &= \frac{y+\sqrt{y^2+4}}{2},
\end{align*}
$\Gamma$ is the standard gamma function and $_{1}{F}_1$ is Kummer's confluent hypergeometric function of the first kind. This is used to calibrate our numerically obtained results. In the geometric optics limit, the images are formed at: 
\begin{equation}\label{eq:img_loc_PML}
    x_{\pm} = \frac{1}{2}\left(y\pm \sqrt{y^2+4}\right),
\end{equation}
where $x_{+}$ is the location of the minima image and $x_{-}$ is the location of the saddle image. The magnifications of these images with respect to the unlensed signal are,
\begin{equation}\label{eq:img_mag_PML}
    \mu_{\pm} = \frac{1}{2}\pm \frac{y^2+2}{2y\sqrt{y^2+4}}~.
\end{equation}
The lensing time delay of the saddle image with respect to the minima image can be found to be,
\begin{equation}\label{eq:time_delay_PML}
    t_{\rm peak} = \frac{1}{2}y\sqrt{y^2+4} + \ln\left(\frac{\sqrt{y^2+4}+y}{\sqrt{y^2+4}-y}\right).
\end{equation}

\subsubsection{Singular Isothermal Sphere}
\label{sec:SIS_amplification_factor}
The singular isothermal sphere (SIS) models a spherically symmetric mass agglomeration with its constituents behaving like an ideal gas in hydrostatic equilibrium. The surface mass density of the SIS is determined by its velocity dispersion $\sigma_v$ and is expressed as \cite{schneider2013gravitational}:
\begin{equation}\label{eq:SIS_sigma}
    \Sigma(\vec{x}) = \frac{\sigma_v^2}{2G\xi_0 x}.
\end{equation}
By choosing:
\begin{equation}
    \xi_0 = \frac{4\pi\sigma_v^2}{c^2}\frac{D_{\rm L}D_{\rm LS}}{D_{\rm S}},
\end{equation}
and using equations \eqref{eq:sigma_to_psi}, \eqref{eq:critical_density} and \eqref{eq:SIS_sigma}, the lensing potential for SIS can be derived as:
\begin{equation}\label{eq:sis_lensing_potential}
    \psi_{\rm SIS}(x) = x \,.
\end{equation}

Unlike the PML, SIS does not have a closed-form analytic solution for $\Ff$. However, there exists a summation form for the amplification factor for SIS \cite{Matsunaga:2006uc},
\begin{align}
    F(w) = & \exp\left(\frac{i}{2}wy^2\right)\sum_{n=0}^{\infty}\frac{\Gamma\left(1+n/2\right)}{n!}\left(2w\right)^{n/2}\nonumber\\
    & \exp\left(\frac{i3\pi n}{4}\right)\hspace{2pt}_{1}{F}_1\left(1 + \frac{n}{2},1;-\frac{i}{2}wy^2\right)~.
    \label{eq:ampfac_SIS_summation}
\end{align}

We make use of Eq.\eqref{eq:ampfac_SIS_summation} to calibrate our numerical results by keeping terms up to $n=500$. For $y<1$, two images form in the geometric optics limit at the locations:
\begin{equation}\label{eq:img_loc_SIS}
    x_\pm = y \pm 1.
\end{equation}
Similar to PML, one of the images (at $x_+$) is a minima and the other (at $x_-$) is a saddle image. The magnifications of the minima and the saddle images with respect to the unlensed waveform are,
\begin{equation}\label{eq:img_mag_SIS}
    \mu_{\pm} = 1\pm\frac{1}{y} \,,
\end{equation}
respectively, for $y<1$. The lensing time delay between the images is,
\begin{equation}\label{eq:time_delay_SIS}
    t_{\rm peak} = 2 y~,
\end{equation}
which is where the logarithmic peak due to the saddle image can be seen. However, for $y>1$, only a single image is present, which is the \emph{weak lensing} regime. In this work, we restrict to the two image parameter space for the SIS lens.

We define the characteristic mass scale of the SIS lens as the mass contained within $\xi_0$, and is given by:
\begin{equation}\label{eq:SIS_lens_mass}
    M_{\rm L} = \frac{4\pi^2\sigma_v^4}{c^2G}\frac{D_{\rm L}D_{\rm LS}}{D_{\rm S}}.
\end{equation}
In the main text, the lens mass of the SIS model refers to the mass defined in Eq. \eqref{eq:SIS_lens_mass} above.

\bibliography{References}

\end{document}

%% file: preamble.tex
\usepackage[T1]{fontenc}
\usepackage[utf8]{inputenc}
\DeclareUnicodeCharacter{2009}{\nobreakspace}
\usepackage{lmodern}
\usepackage{times}
\usepackage[varg]{txfonts}
\usepackage[normalem]{ulem}
\usepackage{verbatim}
\usepackage[dvipsnames, usenames]{xcolor}
\definecolor{linkcolor}{rgb}{0.0,0.3,0.5}
\usepackage[hypertexnames=false, unicode, colorlinks=true, linkcolor=linkcolor,
citecolor=linkcolor, filecolor=linkcolor,urlcolor=linkcolor,
pdfusetitle]{hyperref}
\usepackage[all]{hypcap}
\usepackage{graphicx}
\usepackage{xspace}
\usepackage{amssymb}
\usepackage{bm} 
\usepackage{microtype}
\usepackage[english]{babel}